\documentclass[aps,pra,twocolumn,a4paper,amsmath,amssymb,showpacs,%
floatfix,superscriptaddress]{revtex4}
\usepackage{braket}
\usepackage{graphicx}
\usepackage{dcolumn}
\usepackage{epstopdf}

\newcommand{\Op}[1]{\boldsymbol{\mathsf{\hat{#1}}}}
\newcommand{\gOp}[1]{\boldsymbol{\hat{#1}}}

\def\openone{\leavevmode\hbox{\small1\kern-3.3pt\normalsize1}}

\begin{document}

\title{Steering the optimization pathway in
  the control landscape using constraints} 

\author{Jos\'e P. Palao}
\affiliation{Departamento de F\'isica Fundamental II and IUdEA, 
Universidad de La Laguna, Spain,
La Laguna 38204, Spain}

\author{Daniel M. Reich}
\affiliation{Theoretische Physik,  Universit\"at Kassel,
  Heinrich-Plett-Str. 40, 34132 Kassel, Germany}

\author{Christiane P. Koch}
\affiliation{Theoretische Physik, Universit\"at Kassel,
  Heinrich-Plett-Str. 40, 34132 Kassel, Germany}

\date{\today}

\begin{abstract}
  We show how additional constraints, restricting the spectrum of the
  optimized pulse or confining the system dynamics, can be used to steer
  optimization in quantum control towards distinct solutions.  
  Our examples are multi-photon excitation in atoms and vibrational
  population transfer in molecules. We show that a spectral constraint
  is most effective in enforcing non-resonant two-photon absorption
  pathways in atoms and avoiding unnecessarily broad spectra in Raman
  transitions in molecules. While a constraint restricting the system
  to stay in an allowed subspace is also capable of identifying 
  non-resonant excitation pathways, it does not avoid spurious peaks
  in the pulse spectrum. Both constraints are compatible with
  monotonic convergence but imply different additional numerical costs.
\end{abstract}

\maketitle

\section{Introduction}
\label{sec:intro}

Quantum optimal control utilizes shaped external fields to reach a
desired target in the best possible way~\cite{RiceBook}. It
has been successfully applied 
in a variety of settings, from femtosecond laser spectroscopy to
nuclear magnetic resonance or quantum information processing, see
Ref.~\cite{BrifNJP10} and references therein for a recent overview. 
The enormous success of quantum control, both in optimal control
theory~\cite{SomloiCP93} and adaptive feedback control
experiments~\cite{JudsonPRL92}, has 
been rationalized in terms of the favorable properties of the 
control landscape~\cite{RabitzScience04}. This landscape
visualizes the optimization target as a function of the control
parameters. Optimization corresponds to a search for the maxima or
minima in the landscape. Success of control is explained by broad
peaks that can easily be climbed~\cite{RabitzIRPC07}. Suboptimal peaks, while not
completely excluded~\cite{PechenPRL12}, seem to play no significant
role in actual control applications. The intuitive picture of the
control landscape can not only elucidate search pathways but also help
to find the mechanism underlying an optimized control field. It is
thus not surprising that the theoretical concept has triggered a
number of experimental
investigations~\cite{WollenhauptJMO05,VogtPRA06,BayerJPB08}. 

Any experiment is, however, subject to  constraints such
as finite pulse power, bandwidth,  time or
frequency resolution. These constraints will necessarily make some of
the optimal peaks in the landscape inaccessible and may lead to traps
and saddle points~\cite{MooreJCP12}. 
In order to search for control solutions in optimal control theory
that can be realized in a given experiment, the experimental
constraints should be included as additional costs in the optimization
functional. For example, the system dynamics can be restricted to a
certain subspace in order to block undesired strong field effects or avoid
decoherence~\cite{JosePRA08}. Formulating the additional costs is
not always straightforward. In particular, 
imposing conditions simultaneously on the spectrum and the shape of an
optimized pulse has proven to be
challenging~\cite{KochPRA04,GollubPRL08,LapertPRA09,SchroederNJP09,ReichJMO13}.
Although they exclude part of the ideal control
landscape~\cite{MooreJCP12}, additional constraints are not
necessarily detrimental. They can also be used to actively steer the
optimization pathway toward a particular solution out of several
available ones. This is the subject of our current work. 

We employ spectral constraints, imposing filters on the optimized
spectrum~\cite{ReichJMO13}, and state-dependent constraints,
restricting the system dynamics to a subspace~\cite{JosePRA08}, to
optimize for non-resonant excitation in atoms and vibrational
population transfer in molecules. Unless one employs a two-photon
rotating wave approximation which excludes resonant one-photon
pathways \textit{a priori}, finding non-resonant transitions poses a
notoriously difficult problem in optimal control theory since it
contradicts the condition of minimal power consumption. This is
particularly dissatisfying in view of the many experimental studies of 
non-resonant two-photon absorption for 
$ns$ to $(n+1)s$ transitions in alkali atoms in the
weak~\cite{MeshulachNature98,MeshulachPRA99,PraekeltPRA04},
strong~\cite{TralleroPRA05,TralleroPRL06,TralleroPRA07} and intermediate field 
regime~\cite{ChuntonovPRA08,ChuntonovJPB08,ZoharPRL08,ChuntonovPRA10}. 
To date,  only solutions using one-photon transitions are
found while the experimental result of \textit{non-resonant} two-photon
control~\cite{MeshulachNature98,MeshulachPRA99,PraekeltPRA04,TralleroPRA05,TralleroPRL06,TralleroPRA07,ChuntonovPRA08,ChuntonovJPB08,ZoharPRL08,ChuntonovPRA10} 
could not be reproduced. 
Here we employ optimal control theory using Krotov's
method~\cite{Konnov99,SklarzPRA02,ReichKochJCP12} and impose 
spectral and state-dependent constraints 
to enforce a non-resonant two-photon solution. We then extend our
study to vibrational population transfer. For this example, 
optimal control calculations have also been hampered by an enormous
spectral spread of the field, so much so that the resulting spectral
widths by far exceed experimentally realistic
values~\cite{KochPRA04,NdongKochPRA10}. We show that for both examples
a spectral constraint successfully suppresses all undesired frequency
components. 

The paper is organized as follows: Section~\ref{sec:method} presents a
brief review of Krotov's method for quantum optimal control. Special 
emphasis is placed on how to include additional constraints in  a
way that preserves monotonic convergence. Multi-photon absorption in
sodium atoms is studied in Sec.~\ref{sec:Na} for two different
optimization targets -- maximizing two-photon absorption and
generating a third harmonic with near-infrared light. The problem of
broad spectral widths in vibrational population transfer in molecules
is studied in Sec.~\ref{sec:Rb2}.
Section~\ref{sec:concl} summarizes our findings.

\section{Constraints in Krotov's method}
\label{sec:method}

We briefly review optimization using Krotov's method following
Refs.~\cite{ReichKochJCP12,JosePRA08,ReichJMO13}. 
An optimization problem is defined in terms of the equation of
motion, 
\begin{equation}
  \label{eq:tdse}
  \frac{d}{dt}|\psi(t)\rangle = - \frac{i}{\hbar}\Op H[\epsilon(t)] 
  |\psi(t)\rangle\,,
\end{equation}
and the optimization functional, 
\begin{equation}
  \label{eq:J}
  J[\{\psi_k\}, {\epsilon}] = J_T[\{\psi_k(T)\}]
  +J_a[\epsilon]
  +J_b[\{\psi_k\}] \,,
\end{equation}
 which consists of the target and additional constraints. 
Here $J_T$ is target functional, evaluated at final time $T$, and 
$\{\psi_k(t)\}$ are a set of state vectors which all fulfill
Eq.~\eqref{eq:tdse}. $\epsilon(t)$ represents the control
variable, e.g., the electric field of a laser pulse.
The additional constraints are assumed to depend either on the control 
or on the states, 
\begin{eqnarray}\label{eq:Ja}
  J_a &=&\int_0^T{g_{a}(\epsilon,t) ~ dt } \,, \\
  J_b &=&\int_0^T{g_{b}(\{\psi_k\},t) ~ dt }. \label{eq:Jb}
\end{eqnarray}
We first present the optimization equations for the most general form
of $g_a(\epsilon,t)$ preserving monotonic convergence in
Section~\ref{subsec:control}, followed by a discussion of optimization
under state-dependent constraints 
$g_{b}(\{\psi_k\},t)$ in Section~\ref{subsec:state}. 

\subsection{Spectral constraints}
\label{subsec:control}

We have recently shown that a monotonically converging optimization
algorithm is obtained if the constraint depending on the control is
formulated in terms of a positive semi-definite
kernel~\cite{ReichJMO13}, 
\begin{subequations}
  \begin{eqnarray}
    \label{eq:frequency constraint}
    g_a(\epsilon,t) &=& \frac{1}{2\pi} \int_{0}^{T}
    {\Delta\epsilon(t) K(t-t') \Delta\epsilon(t') ~ dt'} \\
    \label{eq:Kcondition}
    \bar{K}(\omega)&\geq&0 \quad \forall ~ \omega\,,
  \end{eqnarray}
\end{subequations}
where $\bar{K}(\omega)$ is the Fourier transform of $K(t-t')$.
This way, one can enforce constraints which depend both on time and
frequency. 
Since the derivation of the update equation requires
evaluation of $\frac{\partial g_a}{\partial \epsilon}$ as a function of
time \cite{SklarzPRA02,JosePRA03,ReichKochJCP12}, 
the  Fourier transform of $\bar{K}(\omega)$ should have a closed
form in addition to being positive semi-definite. An obvious choice
are Gaussian kernels,
\begin{subequations}
\begin{eqnarray}
  \label{eq:filter}
  \bar{K}(\omega) &=& \lambda^0_a - \sum_j \frac{\lambda_a^j}{2} 
    \left[ e^{-\frac{(\omega-\omega_j)^2}{2\sigma_j^2}} + 
      e^{-\frac{(\omega+\omega_j)^2}{2\sigma_j^2}} \right] \,,\\
    \label{eq:kernel}
    K(t-t') &= & 2 \pi \lambda^0_a \delta(t-t') \\
    &&- \sum_j \lambda_a^j \sqrt{2\pi \sigma_j^2} 
    \cos[\omega_j (t-t')] e^{-\frac{\sigma_j^2 (t-t')^2}{2}} \,,\nonumber
\end{eqnarray}
\end{subequations}
which come with the additional advantage of smoothness which is
desirable in view of numerical stability. 
For (approximately) non-overlapping Gaussians in frequency domain,
monotonic convergence is obtained  if 
\begin{equation}
  \label{eq:condlambda}
  \lambda_a^j \leq 2 \lambda^0_a \quad \forall\,j \neq 0 \,.
\end{equation}
Note that we assume real pulses in Eq.~\eqref{eq:filter} which is why
the kernel is symmetric. An extension to complex pulses is
straightforward by mapping it to a real pulse on a
time grid of twice the size.
The first term in Eq.~\eqref{eq:kernel} reproduces the standard choice
for $g_a$ which minimizes the change in 
intensity~\cite{JosePRA03} with constant shape function. For
$\lambda_a^j>0$ ($\lambda_a^j<0$), the 
kernel~\eqref{eq:kernel} implements a 
frequency pass (filter) for $\Delta\epsilon(t)$
around the frequencies $\omega_j$. 
Due to the condition~\eqref{eq:condlambda}, the strength
of frequency passes that still allow for monotonic convergence is
restricted. This reduces their effectiveness in practice, 
and frequency passes should rather be
enforced by expressing them as a sum over many frequency filters.
An amplitude constraint
with non-constant shape function $S(t)$ can be reintroduced additively in
time domain for $\lambda_a^j < 0$, setting $\lambda^0_a =0$. 
The update equation for the control at iteration $i+1$
for Gaussian band filters and an additional amplitude constraint
imposed by a shape function $\lambda_0 / S(t)$ is obtained as  
\begin{widetext}
  \begin{eqnarray}
    {\epsilon^{(i+1)}(t)} & = & \epsilon^{(i)}(t) + \sum_j
    \frac{\lambda_a^j S(t)}{2 \pi \lambda_0} \sqrt{2\pi \sigma_j^2}
    \int_0^T{\cos[\omega_j (t-t')]\, e^{-\frac{\sigma_j^2
          (t-t')^2}{2}} \left({\epsilon^{(i+1)}(t')} -
        \epsilon^{(i)}(t')\right) ~ dt'} \nonumber \\ 
    & & + \frac{S(t)}{\lambda_0} \mathfrak{Im} \left\{\sum_k
      \Braket{\chi_k^{(i)}(t)|\frac{\partial \Op H}{\partial
          \epsilon}|\psi_k^{(i+1)}(t)}  
      + \frac{1}{2} \sigma(t) \sum_k
      \Braket{\Delta\psi_k(t)|\frac{\partial \Op H}{\partial
          \epsilon}|\psi^{(i+1)}_k(t)} \right\} .\ 
    \label{eq:newupdate}
  \end{eqnarray} 
\end{widetext}
The adjoint states $\{\chi_k(t)\}$ are subject to the same equations of
motion, Eq.~\eqref{eq:tdse}, as the $\{\psi_k(t)\}$ but their
'initial' condition is given at the final time $T$, i.e., they are
propagated backward in time. The specific form of the 'initial'
condition is determined by the final-time target, 
\begin{equation}
  \label{eq:chiT}
  |\chi_k^{(i)}(T)\rangle = -\nabla_{\psi_k} J_T\big|_{\{\psi^{(i)}_k(T)\}} \,,
\end{equation}
which is evaluated using the  forward-propagated states
$\{|\psi^{(i)}_k(T)\rangle\}$. 

In the examples presented below, we assume the interaction with the
control to be linear, 
\begin{equation}
  \label{eq:H}
  \Op H[\epsilon] = \Op H_0 + \gOp\mu\epsilon(t)\, 
\end{equation}
and the target functional to be convex in the states. The
latter allows for the choice $\sigma(t)\equiv 0$, 
and Eq.~\eqref{eq:newupdate} reduces to
\begin{widetext}
  \begin{eqnarray}
    {\epsilon^{(i+1)}(t)} & = & \epsilon^{(i)}(t) +
    \frac{S(t)}{\lambda_0} \mathfrak{Im} \left\{\sum_k
      \Braket{\chi_k^{(i)}(t)|\gOp\mu|\psi_k^{(i+1)}(t)}\right\}  \nonumber \\ 
    & & +\sum_j
    \frac{\lambda_a^j S(t)}{2 \pi \lambda_0} \sqrt{2\pi \sigma_j^2}
    \int_0^T{\cos[\omega_j (t-t')]\, e^{-\frac{\sigma_j^2
          (t-t')^2}{2}} \left({\epsilon^{(j+1)}(t')} -
        \epsilon^{(j)}(t')\right) ~ dt'}  \,,
    \label{eq:update}
  \end{eqnarray} 
\end{widetext}
with the third term due to the spectral constraint. In order to solve 
Eq.~\eqref{eq:update} which is implicit in $\epsilon^{(i+1)}(t)$, we
rewrite it as a Fredholm integral equation of
the second kind in $\Delta \epsilon(t) =
\epsilon^{(i+1)}(t) - \epsilon^{(i)}(t)$, 
\begin{equation}
  \label{eq:Fredholm}
  \Delta \epsilon(t) = I(t) + \gamma \int_0^T{\mathcal{K}(t,t') \Delta
    \epsilon(t') ~ dt'} .\ 
\end{equation}
The inhomogeneity $I(t)$ depends on the unknown states 
$\{\psi_k^{(i+1)}(t)\}$, cf. Eq.~\eqref{eq:update}. We 
approximate them by calculating $\Delta\epsilon(t)$ without frequency 
constraints and solve the Fredholm equation, mapped
from the interval $[0,T]$ to $[0,1]$, using the method of
degenerate kernels with triangularly shaped basis
functions~\cite{Volk79,Volk85}. This corresponds to writing  
$\mathcal{K}_N(t,t')=\sum_{j=1}^N\alpha_j(t)\delta_j(t')$
for an $N$th order approximation with 
\begin{equation}\label{eq:alpha}
  \alpha_{j}\left(t\right)=
\begin{cases}
  1-N\left|t-\frac{j}{N}\right|, & \frac{j-1}{N}\leq t\leq\frac{j+1}{N} \\
  0, & \text{else} \end{cases} \,,
\end{equation}
and solving a system of linear equations, 
\begin{equation}
  \label{eq:lin}  
\left[\openone_{N+1} -\gamma\boldsymbol{\mathsf{C}} \right] 
\vec{X} = \gamma \vec{b} \,,
\end{equation}
with matrix elements
\[
C_{jk} = \sum_{i=0}^{n}K\left(t_{j},t_{i}\right)
\int_{0}^{1}\alpha_{i}\left(t\right)\alpha_{k}\left(t\right)\ dt
\equiv\sum_{i=0}^{n}K\left(t_{j},t_{i}\right)A_{ik} \,,
\]
where 
\[
A_{ik} = \begin{cases}
\frac{1}{3n}, & \text{for }i=k=0\;\text{or}\;i=k=n\\
\frac{2}{3n}, & \text{for }i=k,1\leq i\leq n\\
\frac{1}{6n}, & \text{for }i=k+1\text{ or }i=k-1\\
0, & \text{else}\end{cases}
\]
and 
\[
b_{k} = \int_{0}^{1}I\left(t\right)\left[\sum_{i=0}^{n}K
\left(t_{k},t_{i}\right)\alpha_{i}\left(t\right)\right]\ dt  \,.
\]
The solution to Eq.~\eqref{eq:Fredholm} is then given by
\begin{equation}
  \label{eq:solFred}
  \Delta\epsilon(t) = I(t) + \sum_{j=0}^N X_j \alpha_j(t)
\end{equation}
where $X_j$ is the solution of Eq.~\eqref{eq:lin}. 

To summarize, optimization in the presence of spectral constraints
proceeds in two steps: It requires forward (backward)
propagation of the states $|\psi_k(t)\rangle$ (adjoint states
$|\chi_k(t)\rangle$) according to Eq.~\eqref{eq:tdse} 
and evaluation of the update,
Eq.~\eqref{eq:update}, without the spectral constraint, i.e., 
$\lambda_a^j=0$ for all $j$. This yields the input for $I(t)$ in
Eq.~\eqref{eq:Fredholm} which is solved in the second step.

\subsection{State-dependent constraints}
\label{subsec:state}

State-dependent constraints can be employed to optimize a
time-dependent expectation value or enforce the system to stay within
a subspace of the total Hilbert space~\cite{JosePRA08}. It takes the
form 
\begin{equation}
  \label{eq:g_b}
  g_b\left(\{\psi_k(t)\},t\right) =
  \frac{\lambda_b}{T N} \sum_{k=1}^N
  \langle \psi_k(t)|\Op{D}(t)|\psi_k(t)\rangle\,,
\end{equation}
where the dependence on the states is quadratic.
We will employ in the examples below $\Op D(t)=\Op P_{allow}$. For
this specific choice, $\sigma(t)$ in Eq.~\eqref{eq:newupdate} can be
set to zero, and the  update for the control is again 
given by Eq.~\eqref{eq:update}, possibly with $\lambda_a^j=0$ for all $j$.
Any other choice of $\Op D(t)$ requires non-zero
$\sigma(t)$ and use of Eq.~\eqref{eq:newupdate} as discussed in
Ref.~\cite{ReichKochJCP12}. 
% \begin{equation}
%   \label{eq:sigma}
%   \sigma(t) = \bar{C}(T-t)\,,
% \end{equation}
% where again we assume the final-time target to depend on the states at
% most quadratically (the general case is discussed in
% Ref.~\cite{ReichKochJCP12}). $\bar{C}$ is given by  
% \begin{equation}
%   \label{eq:barC}
%   \bar{C} = 2C-\epsilon_C
% \end{equation}
% with $\epsilon_C$ a non-negative number. $C$ is determined by the
% state-dependent constraint and can be estimated analytically as the
% supremum over the first-order derivatives of $g_b$ or approximated
% numerically using the optimization history~\cite{ReichKochJCP12}. The
% analytical estimate becomes particularly straightforward for a
% quadratic dependence of $g_b$ on the states. It is then given in terms
% of the eigenvalue of $\Op D(t)$ with the largest
% magnitude. For the specific case that $\Op D(t)$ corresponds to a
% projector, the eigenvalue with largest magnitude is one (or zero), and 
% \begin{equation}
%   \label{eq:C}
%   C \le -\frac{\lambda_b}{NT}\,.   
% \end{equation}
% Note that the sign of $g_b$ is crucial, and minimizing the population
% in a forbidden subspace is not equivalent to maximizing the population
% in an allowed subspace. While the former implies a largest eigenvalue
% of one, for the latter the largest eigenvalue is
% zero and allows for $C=0$ and subsequently $\sigma(t)\equiv 0$.  We
% will employ this choice in the examples below, such that the update
% of the control is simply given by Eq.~\eqref{eq:update}. 

A state-dependent constraint does not only affect the update equation
for the control via $\sigma(t)$ but also the equation of motion for
the adjoint states which follows from the first order
extremum condition on the optimization
functional~\cite{JosePRA08}. When evaluating the
derivatives, an additional dependence of the optimization functional
on the states due to $g_b$ yields an additional term in the
equation of motion. The corresponding inhomogeneous Schr\"odinger
equation reads
\begin{equation}
  \label{eq:inhom_tdse}
  \frac{d}{dt}|\chi(t)\rangle =
  -\frac{i}{\hbar} \Op H[\epsilon(t)]|\chi(t)\rangle
  +\lambda_b \Op P_{allow}|\psi(t)\rangle\,.
\end{equation}
It is evaluated for the 'old' control, $\epsilon^{(i)}(t)$, and the 'old'
state, $|\psi^{(i)}(t)\rangle$ and can be solved by a modified Chebychev
propagator~\cite{NdongKochJCP09}. 

To summarize, optimization in the presence of state-dependent
constraints  requires forward 
propagation of the states $|\psi_k(t)\rangle$ according to
Eq.~\eqref{eq:tdse}, solution of an inhomogeneous Schr\"odinger
equation, Eq.~\eqref{eq:inhom_tdse}, for the  
backward propagation of the adjoint states
$|\chi_k(t)\rangle$, and evaluation of the update,
Eq.~\eqref{eq:update}. 

\section{Control of non-resonant two-photon absorption}
\label{sec:Na}

We compare Krotov's method using a spectral constraint and using a
state-dependent constraint to maximize the non-resonant two-photon
absorption in sodium atoms. Our model,
\begin{equation}
  \label{eq:Hatom}
  \Op H[\epsilon] = \sum_j |j\rangle\langle j| + 
  \epsilon(t)\sum_{i\neq j}\mu_{ij}|j\rangle\langle i|\,,
\end{equation}
comprises of the levels
$|j\rangle=|3s\rangle$, $|4s\rangle$ and $|np\rangle$ ($n=3,\ldots,8$) and all
$|ns\rangle\to|n'p\rangle$ dipole-allowed transitions. The energies
and dipole moments are taken from Ref.~\cite{NIST}.  
For the spectral constraint,
forward and backward propagation involve solution of the standard 
time-dependent Schr\"odinger equation, Eq.~\eqref{eq:tdse}, which is
carried out by a Chebychev propagator~\cite{RonnieCheby84}. In contrast, 
for the state-dependent constraint, an inhomogeneous Schr\"odinger equation,
cf. Eq.~\eqref{eq:inhom_tdse}, governs the backward propagation. It
can be solved with a modified Chebychev
propagator~\cite{NdongKochJCP09} which is most efficient when high
accuracy is desired. Here, we simply utilize a zeroth order
approximation as discussed in Ref.~\cite{JosePRA08}: We calculate 
$\exp[i\Op H \Delta t]|\chi(t_i)\rangle$ by diagonalization of the
Hamiltonian with the time dependence  evaluated at $t_i+\Delta t$ and
assumed constant over $\Delta t$. The inhomogeneous term in
Eq.~\eqref{eq:inhom_tdse} is approximated by 
$\lambda_b/2\left(\Op P_{allow}|\psi(t_i)\rangle + 
\Op P_{allow}|\psi(t_{i+1})\rangle\right)$. We use 4096 time grid
points, ensuring a sufficiently small $\Delta t$ for the approximation
to be valid. 
The guess pulse is chosen to be a Gaussian centered around the
two-photon transition frequency, $\omega_{3s,4s}/2$. The shape
function, $S(t)$ in Eq.~\eqref{eq:newupdate}, takes the form $S(t) =
\sin^2(\pi t/T)$. 

We consider two
different targets, to maximize population in $|4s\rangle$ and in 
$\left(|3s\rangle+|7p\rangle\right)/\sqrt{2}$. The $|7p\rangle$ state
is reached from the ground $|3s\rangle$ state by a (2+1) transition
via the $|4s\rangle$ state using near-infrared
photons~\cite{TralleroPRA07,GandmanPRA07,ClowPRL08}. The target 
$\left(|3s\rangle+|7p\rangle\right)/\sqrt{2}$ yields a maximum
transition dipole between $|3s\rangle$ and $|7p\rangle$ and 
thus corresponds to maximizing harmonic generation of ultraviolet
light~\cite{RybakOE08}.  

\begin{figure}[tb]
  \centering
  \includegraphics[width=0.9\linewidth]{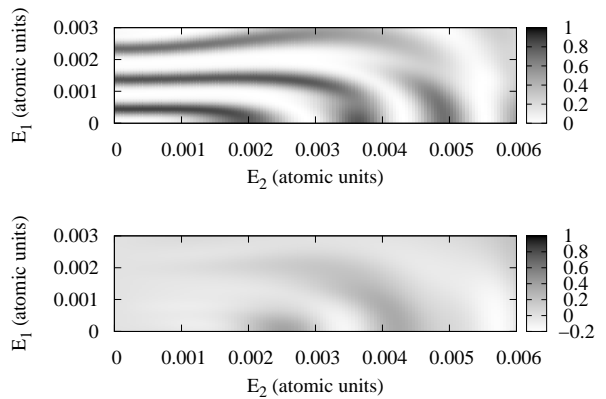}
  \caption{
    Transition probability landscape for non-resonant two-photon
    absorption ($\langle 4s|4s\rangle$, top) and harmonic
    generation ($2\mathfrak{Re}\left[\langle 7p|3s\rangle\right]$,
    bottom) for transform-limited $50\,$fs Gaussian pulses,
    parametrized by Eq.~\eqref{eq:Eparam}. 
  }
  \label{fig:landscape}
\end{figure}
The constraints are necessary since two obvious control strategies are
available -- resonant two-color 
one-photon transitions with frequencies $\omega_{3s,3p}$ and
$\omega_{3p,4s}$  or off-resonant two-photon transitions with
frequencies close to $\omega_{3s,4s}/2$.
This is illustrated by Fig.~\ref{fig:landscape} which displays the two 
figures of merit, population of $|4s\rangle$ and maximum coherence on
the $|3s\rangle\to|7p\rangle$ transition at the end of the pulse, as a
function of one-photon and two-photon amplitudes. 
The visualization of the control landscape
is based on parametrizing the field by
\begin{eqnarray}
  \label{eq:Eparam}
  E(t) &=& e^{-\frac{(t-T)^2}{2\tau^2}}\big\{
    E_1\left[\cos\left(\omega_{3s,3p}t\right)
      +\cos\left(\omega_{3p,4s}t\right)\right]\nonumber \\
    && \quad\quad\quad\quad+ E_2 \cos\left(\omega_{3s,4s}t/2\right)\big\}  \,.
\end{eqnarray}
The two different solutions, resonant one-photon transitions and
non-resonant two-photon transitions, are clearly visible in the upper
panel of Fig.~\ref{fig:landscape}. A possible solution to achieving 
maximum population in $|4s\rangle$ is a two-photon
$\pi$-pulse~\cite{ShoreBook11}. For one-photon transitions, this
requires equal Rabi frequencies on both
transitions~\cite{ShoreBook11}. Since the transition dipole moments
for the $|3s\rangle\to|3p\rangle$ and the $|3p\rangle\to|4s\rangle$
transitions are fairly similar, this condition can almost be fulfilled
even by identical $E_1$ on both transitions as assumed in
Eq.~\eqref{eq:Eparam}. Correspondingly, a series of dark shaded
regions is found in Fig.~\ref{fig:landscape} for $E_2=0$,
for a two-photon $\pi$-pulse, $3\pi$-pulse and $5\pi$-pulse. The
population of $|4s\rangle$ becomes smaller as $E_1$ is
increased. This is due to the dynamic Stark shift getting
larger and shifting the transition off resonance. Analogously to the
series of dark shaded regions as a function of $E_1$ for $E_2=0$, a
similar series is found as a function of $E_2$ for $E_1=0$.
The amplitude for a non-resonant two-photon
$\pi$-pulse with a duration of $50\,$fs corresponds to
$E_2=0.00201\,$a.u. Since our 
parametrization allows only for transform-limited pulses, population
transfer is not complete at this value of $E_2$. This is again due to
the large dynamic Stark shift. It can be compensated by chirping the
pulse~\cite{TralleroPRA05} but for the sake of a simple pulse
parametrization that allows for visualizing the control landscape in
terms of two parameters, we
only analyze transform-limited pulses. Another reason for
the population in $|4s\rangle$ to be smaller than one is population
leakage to  $|7p\rangle$ since the transition energy
$\omega_{4s,7p}$ is very close to $\omega_{3s,4s}/2$. 
When both $E_1$ and $E_2$ are non-zero, the purely one-photon and
purely two-photon solutions are smoothly connected by pulses which
contain both spectral components. 
The lower panel of Fig.~\ref{fig:landscape} illustrates that the
solution to maximizing coherence on the $|3s\rangle\to|7p\rangle$
transition is less obvious. 

\begin{figure}[tb]
  \centering
  \includegraphics[width=0.99\linewidth]{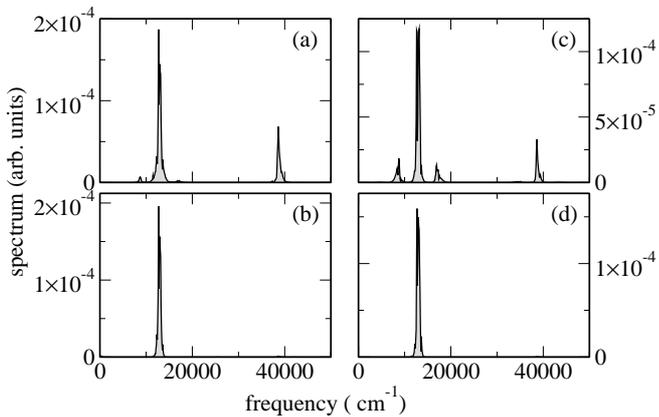}
  \caption{
    Control of non-resonant two-photon absorption: Optimized spectra
    with (b,d) and without (a,c) bandwidth constraint
    ($\lambda_0=400$ left, $\lambda_0=1000$ right). The bandwidth
    constraint consists of filters at $\omega_{3s,3p}$,
    $\omega_{3p,4s}$, $3\omega_L$ and $5\omega_L$ (where
    $\omega_L=\omega_{3s,4s}/2$) with $\sigma_i=0.004\,$a.u. and
    $\lambda_a^i=10^6$ $\forall i$. } 
  \label{fig:spectrum_TPA}
\end{figure}
\begin{figure}[tb]
  \centering
  \includegraphics[width=0.9\linewidth]{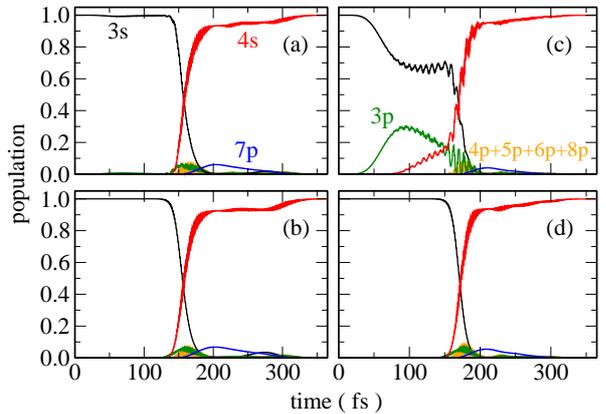}
  \caption{(color online)
    Control of non-resonant two-photon absorption: Population dynamics
    under the optimized fields     with (b,d) and without (a,c) bandwidth
    constraint ($\lambda_0=400$ left, $\lambda_0=1000$ right) as shown in
    Fig.~\ref{fig:spectrum_TPA}.}
  \label{fig:pop_TPA}
\end{figure}
We discuss now how the spectral constraints can be used to steer the
optimization pathway in the control landscape shown in the upper panel
of Fig.~\ref{fig:landscape}. Our guess pulse is of
the form~\eqref{eq:Eparam} with $E_1=0$, $E_2=0.0005\,$a.u., and a pulse 
duration of 50$\,$fs. For this guess pulse, there are two possible
pathways: increasing the intensity to obtain a two-photon
solution, i.e., moving along the $E_2$-axis, or adding new frequencies
to the pulse to obtain the resonant $|3s\rangle\to|3p\rangle$ and
$|3p\rangle\to|4s\rangle$ transitions, i.e., moving along the
$E_1$-axis. Once the guess pulse is fixed, the only free
parameter in the standard Krotov method is $\lambda_0$ which
determines the step size for changes in the control,
cf. Eq.~\eqref{eq:update}. 
For this simple example, it turns out that the choice of $\lambda_0$
is sufficient to steer the optimization pathway in one of the two
possible directions, cf. Fig.~\ref{fig:spectrum_TPA}: A small value of
$\lambda_0$ allows for finding the two-photon solution, i.e., no
peaks at the one-photon frequencies, $\omega_{3s,3p}=16956\,$cm$^{-1}$ and
$\omega_{3p,4s}=8766\,$cm$^{-1}$, are observed in 
Fig.~\ref{fig:spectrum_TPA}(a), whereas for a large value of
$\lambda_0$, these peaks are present,
cf. Fig.~\ref{fig:spectrum_TPA}(c). Figure~\ref{fig:pop_TPA}
displaying the population dynamics under the optimized fields of
Fig.~\ref{fig:spectrum_TPA} confirms this interpretation: The
$|3p\rangle$ state is populated significantly at intermediate times in
Fig.~\ref{fig:pop_TPA}(c). A small value of $\lambda_0$ is, however,
not very useful in general since it often leads to 'exotic' solutions
with spurious peaks at harmonics of the laser frequency such as the peaks
at $3\omega_L$ and $5\omega_L$ in Fig.~\ref{fig:spectrum_TPA} (the
latter being outside of the figure's scale). In our simple example,
these peaks can simply be removed from the spectrum without
compromising the figure of merit or affecting the population
dynamics. However, as we show below, this is not always the case. The
solution is then an 
increase in $\lambda_0$ but this implies that any capability of
steering the optimization pathway in the algorithm without additional
constraints is lost. 

The situation changes when the spectral constraint is included in the
optimization functional, cf. Fig.~\ref{fig:spectrum_TPA}(b) and
(d). No matter what is the value of 
$\lambda_0$, a pure two-photon solution is found. Additionally, the
spurious peaks at the higher harmonics can be suppressed by adding a
filter at the corresponding frequencies. The non-resonant character of
the excitation is confirmed by Fig.~\ref{fig:pop_TPA}(b) and (d) where
almost no population in $|3p\rangle$ is observed. Moreover, the
population leakage to higher $|p\rangle$-states is slightly smaller in  
Fig.~\ref{fig:pop_TPA}(b,d) than in Fig.~\ref{fig:pop_TPA}(a,b).

\begin{figure}[tbp]
  \centering
  \includegraphics[width=0.95\linewidth]{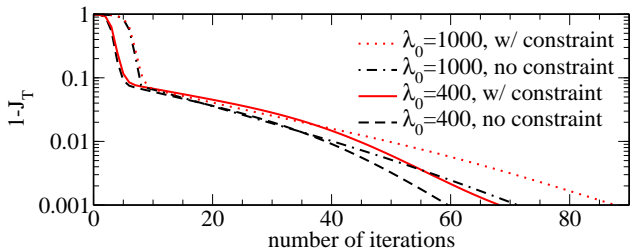}
  \caption{(color online) Convergence toward the optimum 
    with and without spectral
    constraint for two-photon absorption and two different
    optimization step sizes. 
  }
  \label{fig:conv_Na}
\end{figure}
The enhanced functionality of Krotov's method including spectral
constraints comes at a price. This is illustrated by
Fig.~\ref{fig:conv_Na} which shows how the final-time target $J_T$
functional approaches its optimum, $J_T=1$. Independently of the value
of $\lambda_0$, more iterations are required when the 
spectral constraint, which makes the control problem harder, is
included. However, the increase in the number of iterations is very
moderate. The actual additional computational cost due to the spectral
constraint depends on the complexity of the quantum system. In the
current example, the forward and backward propagation are numerically
very inexpensive. The solution of the Fredholm equation,
Eq.~\eqref{eq:Fredholm}, then represents
a significant computational overhead~\cite{ReichJMO13}. However, for
complex quantum systems, propagation of the states and adjoint
states requires by far most of the numerical effort, and the
additional cost of solving the Fredholm equation becomes negligible. 
Figure~\ref{fig:conv_Na} also shows a faster convergence for a smaller
value of $\lambda_0$. This is not surprising since a smaller value of
$\lambda_0$ implies a larger change in the control, 
cf. Eq.~\eqref{eq:update}. 

\begin{figure}[tbp]
  \centering
  \includegraphics[width=0.99\linewidth]{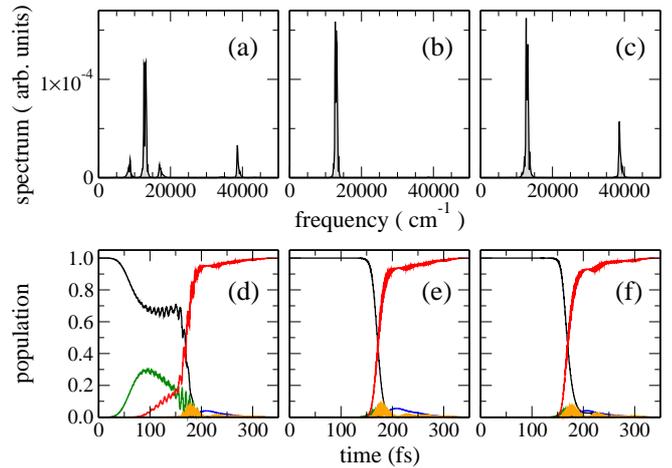}
  \caption{(color online) Control of two-photon absorption: Spectra
    (upper panel) and
    population dynamics (lower panel) for optimization with spectral
    (middle), state-dependent (right) and no constraint (left) for a 
    50$\,$fs guess pulse with $E_2=0.0005\,$a.u., 
    $\lambda_0=1000$, and  $\lambda_bT=-1$. 
    The same filters and $\lambda_a^i$ for the spectral constraint 
    as in Fig.~\ref{fig:spectrum_TPA} and the same color coding for
    the population dynamics as in Fig.~\ref{fig:pop_TPA} are used.
  } 
  \label{fig:compareconst_TPA}
\end{figure}
The control strategy using resonant two-color one-photon
transitions populates the $|3p\rangle$ state,
cf. Fig.~\ref{fig:pop_TPA}(c). Alternatively to
employing a spectral constraint, it should therefore be possible to
enforce a non-resonant two-photon solution with a state-dependent
constraint that suppresses the population of $|3p\rangle$ at any
time. To this end, we define the allowed subspace to be spanned by 
$|3s\rangle$ and $|4s\rangle$ and maximize population in this subspace
for all times using  a state-dependent constraint.
Figure~\ref{fig:compareconst_TPA} compares optimization of two-photon
absorption without any additional constraint (a,d) to that with the
spectral constraint used before (b,e) and the state-dependent
constraint just defined (c,f). The peak amplitude of the initial
Gaussian guess pulse corresponds to a two-photon $\pi/4$-pulse. Both
optimizations with an additional constraint avoid population of the
$|3p\rangle$ state completely, cf. the green lines in the lower part
of Fig.~\ref{fig:compareconst_TPA}. Correspondingly, 
the one-photon peaks at $\omega_{3s,3p}=16956\,$cm$^{-1}$ and
$\omega_{3p,4s}=8766\,$cm$^{-1}$ are missing in the spectrum obtained
with the state-dependent constraint in 
Fig.~\ref{fig:compareconst_TPA}(c). However, only the spectrum 
obtained with the spectral constraint in 
Fig.~\ref{fig:compareconst_TPA}(b) corresponds to a pure two-photon
solution. This observation emphasizes that one should use a
mathematical formulation of the constraint that best captures the
physical goal, in our case, the non-resonant two-photon solution.

\begin{figure}[tbp]
  \centering
  \includegraphics[width=0.99\linewidth]{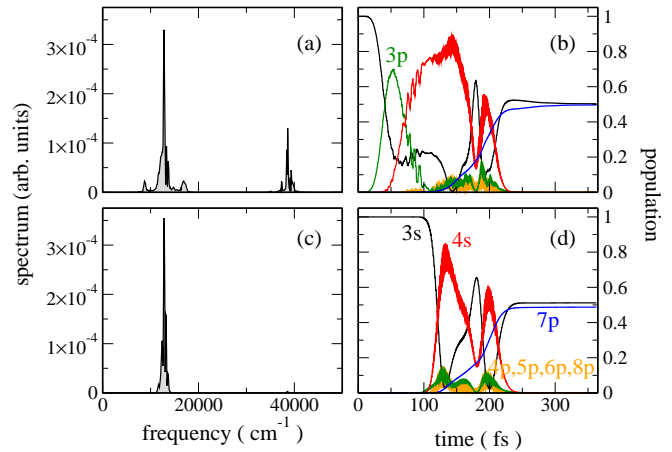}
  \caption{(color online) Control of harmonic generation: Spectrum and
    population dynamics for optimization with (lower panel) 
    and without (upper panel) spectral  constraint for
    $\lambda_0=400$. 
    The same filters and $\lambda_a^i$ as in Fig.~\ref{fig:spectrum_TPA}
    are used.}
  \label{fig:HHG}
\end{figure}
Maximizing the $|3s\rangle\to|7p\rangle$ transition dipole represents
a somewhat harder optimization problem than maximizing two-photon
absorption, and transform-limited pulses are not sufficient to
approach the optimum, cf. the lower panel of
Fig.~\ref{fig:landscape}. The difficulty of the optimization problem
is reflected in the fact that  optimization without any
additional constraint always yields spectra that contain the
one-photon peaks at $\omega_{3s,3p}=16956\,$cm$^{-1}$ and
$\omega_{3p,4s}=8766\,$cm$^{-1}$, cf. the upper panel of
Fig.~\ref{fig:HHG}. This is true even for very large values of
$\lambda_0$, up to 100000, that imply a very cautious search in small
steps. The one-photon character of the transition is confirmed by the
large population of $|3p\rangle$, up to 70\% at about $t=50\,$fs,
in Fig.~\ref{fig:HHG}(b). 
In addition to the two-photon and one-photon peaks, also a peak at
$3\omega_L$ is observed in the upper panel of Fig.~\ref{fig:HHG}. This
spectral component is spurious with little influence on the population
dynamics. The broad spectrum of Fig.~\ref{fig:HHG}(a) is in 
contrast to that obtained by optimization under the spectral
constraint which yields a perfect non-resonant two-photon solution,
cf. Fig.~\ref{fig:HHG}(c), demonstrating the effectiveness of the
spectral constraint. In both cases, the $|3s\rangle$ state is
completely depleted and later refilled, cf. the black lines in
Fig.~\ref{fig:HHG}(b) and (d). 

\begin{figure}[tbp]
  \centering
  \includegraphics[width=0.9\linewidth]{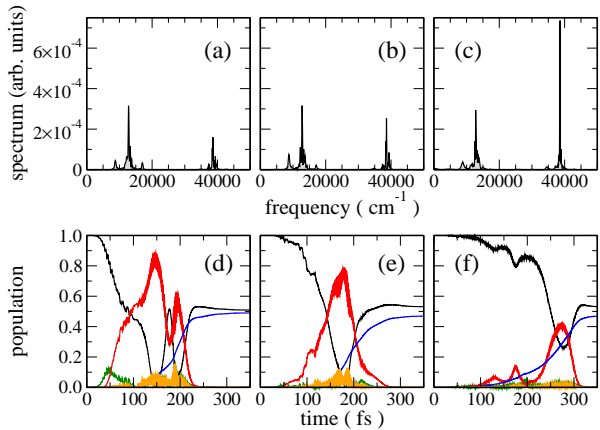}
  \caption{(color online) Control of harmonic generation with a
    state-dependent constraint: Spectra (upper panel) and
    population dynamics (lower panel) for optimization with 
    $\lambda_bT=-0.5$ (left),
    $\lambda_bT=-1.0$ (middle) and $\lambda_bT=-1.5$ (right) 
    and a 50$\,$fs guess pulse with     $E_2=0.000201\,$a.u.,
    $\lambda_0=400$.
  }
  \label{fig:statedep_HHG}
\end{figure}
The effect of a state-dependent constraint is studied in
Fig.~\ref{fig:statedep_HHG} for increasing weight of the constraint,
$\lambda_bT$. The allowed subspace is now defined as
$\{|3s\rangle,|4s\rangle,|7p\rangle\}$. As indicated by the very
different population dynamics observed in
Fig.~\ref{fig:statedep_HHG}(d), (e) and (f), the optimization
identifies very different solutions when changing the weight of
the constraint. However, the corresponding spectra are very complex,
i.e., none of these solutions resembles the simple spectrum obtained
by optimization with the spectral constraint,
cf. Fig.~\ref{fig:HHG}(c). Increasing the weight $\lambda_bT$ leads to
larger widths of each of the spectral peaks with  the optimized
spectrum for the largest value of $\lambda_bT$ containing 
also a peak at $5\omega_L$ (not shown on the scale of
Fig.~\ref{fig:statedep_HHG}(f)). Although the two-photon
peak is central for the population dynamics, cf. the red lines in the
lower part of Fig.~\ref{fig:statedep_HHG}, the contribution of the
additional peaks is needed to realize the desired population transfer. 

\begin{figure}[tb]
  \centering
  \includegraphics[width=0.95\linewidth]{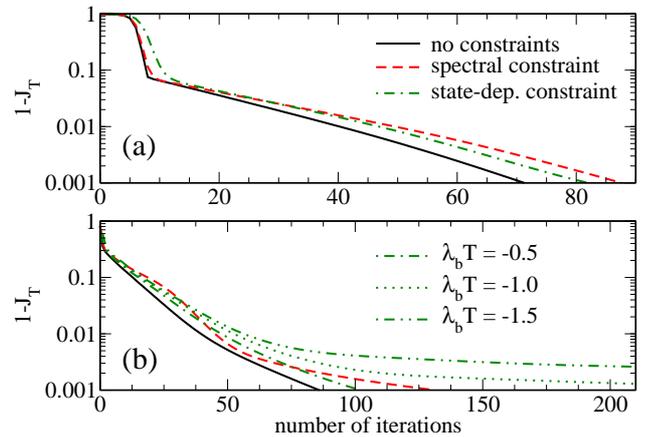}
  \caption{(color online) Convergence toward the optimum 
    with spectral constraint, state-dependent constraint and
    no constraint for two-photon absorption (a) and harmonic
    generation (b). The parameters in (a) correspond to those of
    Fig.~\ref{fig:compareconst_TPA}, in (b) to those of
    Fig.~\ref{fig:statedep_HHG}.
  }
  \label{fig:conv_compareconstraints}
\end{figure}
The convergence behavior of the optimization algorithm  for maximizing
two-photon absorption (upper panel) and maximizing
the transition dipole of the $|3s\rangle\to|7p\rangle$ transition
(lower panel) is shown in Fig.~\ref{fig:conv_compareconstraints},
comparing spectral (red dashed line), state-dependent (green dotted
and dash-dotted lines) and no constraint (black solid line). 
Not surprisingly, restricting the search by additional constraints
increases the number of iterations to reach a prespecified value of
the target functional.  Which of the constraints,
spectral or state-dependent, requires more iterations depends on the
weights $\lambda_a^i$ and $\lambda_bT$. 
The dotted and double-dot-dashed green curves in
Fig.~\ref{fig:conv_compareconstraints}(b) reach $1-J_T=10^{-3}$ after
347, resp. 3146, iterations. This illustrates that too large a value
of the weight can lead the algorithm to get stuck. For both
constraints, the additional numerical effort is not only due to a
larger number of iterations. While the Fredholm equation,
Eq.~\eqref{eq:Fredholm}, needs to be solved for the spectral
constraint as discussed above, the state-dependent constraint requires
backward propagation with an inhomogeneous Schr\"odinger equation,
cf. Eq.~\eqref{eq:inhom_tdse}. Since the latter requires more
applications of the Hamiltonian than propagation for a regular
Schr\"odinger equation~\cite{NdongKochJCP09}, 
the numerical effort due to the inhomogeneity increases with system
complexity. This is in contrast to the spectral constraint where the
additional effort due to the constraint is independent of the system
complexity and depends only on the number of points used in the time
discretization. This represents another important advantage of the
spectral constraint approach.   

\section{Control of vibrational population transfer}
\label{sec:Rb2}

We apply Krotov's method using a spectral constraint to a second
example, vibrational population transfer in Rb$_2$
molecules. Our model accounts for the 32 lowest
vibrational levels in each of two electronic states, the
$X^1\Sigma_g^+$ ground  state and the  
$(1)^1\Sigma_u^+$  electronically excited state. The details of the model
are found in Ref.~\cite{JosePRA08}. 
The time-dependent Schr\"odinger equation for the forward and
backward propagation, given by Eq.~\eqref{eq:tdse}, is solved
using a Chebychev propagator~\cite{RonnieCheby84} and 16384 time grid
points.  
The guess pulse is chosen to be a Gaussian centered around the
frequency of the $X^1\Sigma_g^+(v=0) \to (1)^1\Sigma_u^+(v'=10)$
transition, 
and the shape function is the same as in Sec.~\ref{sec:Na}. 

The optimization goal
consists in driving population from $v=10$ to $v=0$, both in the
electronic ground state, using Raman transitions via the
electronically excited state. 
This type of population transfer is known to yield optimized pulses
with very broad spectra~\cite{KochPRA04,NdongKochPRA10}. 
We therefore apply a spectral constraint to see whether solutions with
more favorable spectra exist and can be identified. Obviously, a
state-dependent constraint 
is of no use in this context, since the many spectral components are
not easily connected to specific levels that could then be assigned 
to the forbidden subspace.

\begin{figure}[tbp]
  \centering
  \includegraphics[width=0.99\linewidth]{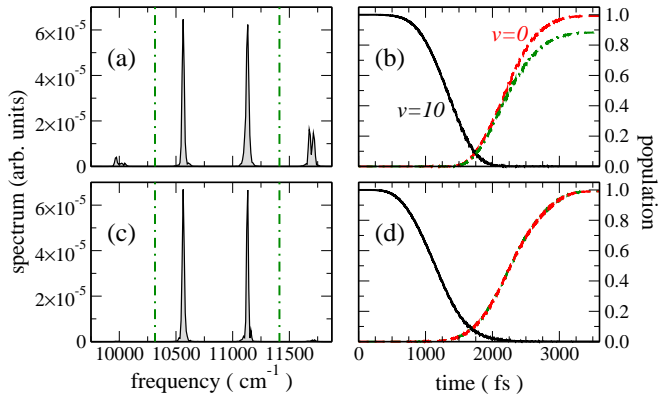}  
  \caption{(color online) 
    Spectra (left) and population dynamics (right)
    for vibrational Raman transfer in
    Rb$_2$ molecules from $v=10$ to $v=0$ optimized with (bottom) and
    without (top) spectral constraint using $\lambda_0=1000$.
    The bandwidth constraint consists of filters at $9440\,$cm$^{-1}$,
    $10000\,$cm$^{-1}$ and $11676\,$cm$^{-1}$
    with $\sigma_i=220\,$cm$^{-1}$ and $\lambda_a^i=10^{5}$ $\forall i$.  
    The green dot-dashed lines in (b,d) represent the population in
    $v=0$ for pulses 
    where all spectral amplitude outside the interval indicated by the
    vertical green lines in (a,c)    was removed.
  }
  \label{fig:Rb2}
\end{figure}
The results of optimization with and without spectral constraint are
shown in Fig.~\ref{fig:Rb2} for 
a Gaussian guess pulse with central frequency
$\omega_L=11127\,$cm$^{-1}$, corresponding to the transition frequency
$\omega_{v=0,v'=10}$, peak amplitude $E_0=10^{-4}\,$a.u. and
pulse duration of 960$\,$fs. In addition to the peak of the guess
pulse and an obvious peak at $\omega_{v=10,v'=10}=10565\,$cm$^{-1}$, 
the spectrum obtained by optimization without constraint contains
peaks at $9440\,$cm$^{-1}$, $10000\,$cm$^{-1}$ and $11676\,$cm$^{-1}$,
cf. Fig.~\ref{fig:Rb2}(a). 
These peaks are not spurious: When removed from the pulse, the
population in the target level $v = 0$ is reduced by more than 10\%,
cf. the red dashed and green dot-dashed lines in
Fig.~\ref{fig:Rb2}(b). 
When using the new algorithm with filters at those frequencies, the
spectral amplitudes are largely reduced. 
Their influence on the population dynamics is negligible, as seen by
the  red dashed and green dot-dashed lines in
Fig.~\ref{fig:Rb2}(d) which are nearly indistinguishable.

This example demonstrates effectiveness of the spectral constraint
for a system which is too complex to guess a simple solution
to the control problem. Indeed, optimization with and without spectral
constraint yields distinct solutions with different spectral
properties. 

\section{Conclusions}
\label{sec:concl}

We have shown how additional constraints can be used in quantum
optimal control to steer the optimization pathway towards one desired
solution out of several possible ones. We have considered non-resonant
excitation of atoms and vibrational Raman transfer in molecules. 
In order to enforce non-resonant absorption, both a spectral
constraint and a state-dependent constraint are effective in
suppressing resonant excitation pathways. However, only the spectral
constraint yields simple spectra without spurious peaks. For
vibrational population transfer using Raman transitions, the spectral 
constraint allows for finding solutions with minimal spectral
support. This is in contrast to unconstrained optimization which
yields spectra consisting of several peaks that are all relevant for
reaching the final-time target. There also exist control problems
where the state-dependent constraint represents the best suited
approach, for example when avoiding population transfer
to states that are resonant with the main pulse
frequencies~\cite{JosePRA08}. In this case, the spectral constraint
would not be helpful. In all of these examples, the additional
constraint allows for identifying different control strategies
than obtained by unconstrained optimization. A similar conclusion is
reached by a related investigation on the control of molecular
orientation using state-dependent and time
contraints~\cite{NdongJMO13}. 

Both constraints imply a larger numerical cost than the standard
optimization without additional constraints. They lead to a moderate 
increase in the number of iterations required to reach a
prespecified value of the final-time target. This reflects that a
constrained control problem is harder to solve. Moreover, the spectral
constraint requires solution of an implicit integral equation for the
change in the control, whereas an inhomogeneous Schr\"odinger equation
needs to be solved when using the
state-dependent constraint. Notably, the additional numerical effort
for the spectral constraint is independent of system size and depends
only on the number of points used in the time discretization. 

In summary, most quantum control problems admit many solutions. In
order to select the 'best' solution, it is crucial to employ a
mathematical formulation of additional constraints that closely
captures the physical desiderata. Spectral constraints
represent a particularly important class of constraints since the
pulse bandwidth in any experiment is necessarily finite. Moreover
smooth spectra with minimal support are typically associated with more
robust solutions. A possible connection between spectral constraints
and robustness of the control will be the subject of future work. 

\begin{acknowledgments}
  We would like to thank Ronnie Kosloff for many valuable
  discussions. DMR and CPK enjoyed hospitality of the Kavli Institute
  of Theoretical Physics. 
  Financial support by the Spanish MICINN (Grant
  No. FIS2010-19998), the Deutsche
  Forschungsgemeinschaft (grant No. KO 2301/2) and in part by the
  National Science Foundation (grant No. NSF PHY11-25915)
  is gratefully acknowledged.
\end{acknowledgments}

%\bibliography{oct}

\end{document}